\documentclass[fleqn,10pt]{wlscirep}
\usepackage[utf8]{inputenc}
\usepackage[T1]{fontenc}
\title{
Mott and Wiedemann Franz Law for Monolayer Graphene for Different Scattering Mechanisms 
}

\author[1]{Purnendu Ray}
\author[1,*]{Kingshuk Sarkar}
\affil[1]{Department of Physics, VIT-AP University Inavolu, Beside AP Secretariat,       Amaravati AP, 522237 India.}
\affil[*]{kingshuk.sarkar@vitap.ac.in}

\begin{document}

\begin{abstract}
In this study, we conducted a comprehensive review and analysis of the thermoelectric responses exhibited by monolayer pristine graphene in response to temperature variations. Employing the Boltzmann transport theory, we rigorously examined and evaluated various thermoelectric coefficients, with particular emphasis on elucidating their behavior under different scattering mechanisms. We derived the analytical expressions for electrical conductivity, thermopower and thermal conductivity at low temperature by Sommerfeld expansion of the Fermi integral. We demonstrated that our numerically obtained values are consistent with the analytical calculations at low temperatures and hence obeying Mott and Wiedeman Franz law. However, the deviation was observed at higher temperatures. Furthermore, we performed theoretical calculations of chemical potential at both low and high temperatures and compared them with our numerically evaluated results at all temperatures. Through extensive calculations and meticulous evaluation, our study contributes to a deeper understanding of the intricate thermoelectric properties inherent in monolayer pristine graphene. 
\end{abstract}
\flushbottom
\maketitle
%

\section*{Introduction}\label{intro}
The study of thermoelectricity has been an intense area of research for a wide variety of metals and semiconductors after the first discovery by German physicist Thomas Johann Seebeck (German pronunciation: zebeck \cite{Ref1} in 1823, hence goes by the name Seebeck effect. Thermoelectricity means generating electricity by applying a temperature difference in a conducting sample, which Seebeck found in thermocouple junctions created by intersections of any pair of metals. The voltage difference (resulting electricity) was generated due to the difference in temperature between the junctions. A decade later, in 1834, the opposite effect was observed by a French physicist,
\begin{figure}[ht]\centering  \includegraphics[width=0.50\linewidth]{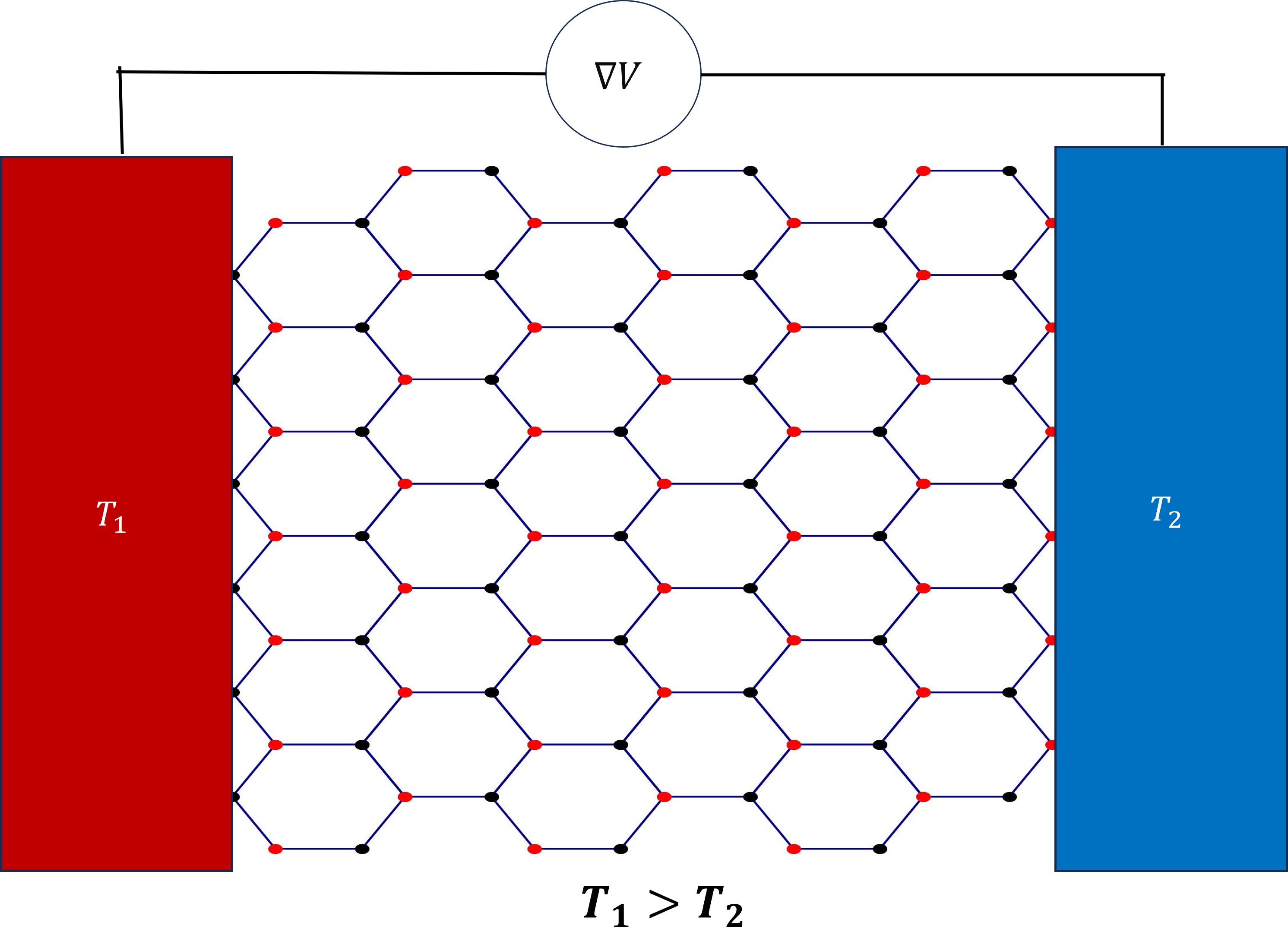}\caption{A schematic representation of thermally induced voltage on a monolayer graphene by two reservoirs having different temperatures. The left reservoir (red) is hotter ( $T_1>T_2$) than the right reservoir (blue). }\label{fig1}\end{figure}
Jean Charles Athanase Peltier where voltage difference applied across a thermocouple was the cause of temperature difference. These two effects are sometimes termed in the literature as the Seebeck-Peltier effect. In the Seebeck effect, under open circuit conditions, the ratio between the measured voltage difference $\nabla V$ and the temperature difference $\nabla T$ is defined as a Seebeck coefficient or sometimes called thermopower, and denoted by $S=-\frac{\mathrm{\nabla V}}{\mathrm{\nabla T}}$ in Fig. (\ref{fig1}). The SI unit of the thermopower is $V/K$, but in the literature, the unit $\mu V/K$ is widely used.   For a particle-hole symmetric system, the thermopower turns out to be zero and hence it can be used to detect the particle-hole asymmetry of a system. This feature can further be utilized to elucidate the details of the electronic structure of a material and the presence of the ambipolar behavior of its constituent charged particle of the material. The other transport coefficients like electrical ($\sigma$) and thermal conductivity ($\kappa$) do not provide those characteristics directly while they provide other valuable information of a material. In the low-temperature regime, thermopower ($S$) and electrical conductivity ($\sigma$) can be connected by Mott's relation \cite{Ref2,Ref3} in the following expression,
\begin{equation}
S=-\frac{\pi^2}{3}\frac{k_B}{e}\frac{T}{T_F}\frac{1}{\sigma\left(\mu\right)}\left[\frac{\partial\sigma}{\partial E}\right]_{E=\mu} \label{eq1}\end{equation}
 This relation is particularly applicable to regular metals and semiconductors due to their high Fermi temperatures $(T_F)$ and it is derived using Sommerfeld expansion for $T\ll T_F$.
Recently, the study of thermoelectricity in graphene and other Dirac materials has gained significant attention in the scientific community, both experimentally\cite{Ref4,Ref5} and theoretically\cite{Ref6, Ref7, Ref8}  . Graphene is a two-dimensional material that takes the form of a honeycomb lattice of $\mathrm{sp}^2$ bonded carbon atoms. It shows outstanding electronic and thermal properties and the electronic band structure is unique due to the relativistic nature of the electron. The valence and conduction band of graphene meet at six vertices of the hexagonal Brillouin zone and form linearly dispersing Dirac cones and zero band-gap at the charge neutrality point\cite{Ref9, Ref10, Ref11, Ref12, Ref13} resulting in a semimetal nature. The electron mobility of graphene can be orders of magnitude higher than the other two-dimensional thermoelectric materials due to weak electron-phonon interaction up to room temperature and it also exhibits large thermal conductivity. The typical Seebeck coefficient of monolayer graphene is measured around $80~ \text{to} ~100 ~ \mu V/K$ in room temperature in Ref. [\cite{Ref5}]. Mott’s formula was observed to be valid for low temperatures from 15-200 K by plugging in the measured conductivity data in Eq. (\ref{eq1}). Later in recent experiment\cite{Ref6}, enhanced thermopower was observed in low disorder graphene at higher temperature and attributed to the strong inelastic scattering due to the charge carriers. Experimentally, a change in the sign of thermopower is observed around the charge neutrality point (CNP) as the majority of carriers transition from electrons to holes. Away from the CNP, thermopower is inversely proportional to the square root of carrier density. 
\par
To determine, the temperature dependence of a material's thermoelectric properties, the Fermi temperature ($T_F$ must remain constant, $T_F$ is related to the carrier density (n) by: $T_F\propto\sqrt n$. For graphene, the carrier density is typically on the order of ${10}^{10} cm^{-2}$, giving a $T_F$ of approximately $1000 K$. This high $T_F$ supports graphene's potential for high-temperature thermoelectric applications. An important parameter in calculating thermoelectric properties is the chemical potential ($\mu$) deriving a general expression for $\mu$ as a function of temperature is challenging, though asymptotic relations can be obtained using the Sommerfeld  expansion.  \cite{Ref14, Ref15, Ref16}  \par 
 While theoretical calculations often consider screened charge impurity scattering as the dominant mechanism, especially for graphene on SiO2 substrates. Whether, for suspended graphene in vacuum the scattering due to substrate impurity is absent. The primary source of scattering in suspended graphene is out-of-phase flexural phonon.  A comprehensive approach to the energy dependence of scattering time remains unexplored. \par
Along with the study of thermopower other transport coefficients like electrical and thermal conductivity are also explored recently due to their intriguing transport phenomena. Since the discovery of graphene electrical and thermal conductivity of graphene has been widely studied both from experimental  and theoretical  aspects. Typically, the electrical conductivity of graphene has a value of $2.3 \ \text{ to} \ 14.6 ~ S/m$.  The experimental value of electronic thermal conductivity is 500-1000 W/m K  at 300K. Both $\kappa_{tot}$ and $\sigma$ show monotonically increasing behavior with temperature.\cite{Ref17,Ref18,Ref19,Ref20}.  In both these carrier transport coefficients the majority contribution comes from phonon. The first attempt to establish a relationship between $\kappa_e$ and $\sigma$ was made by Wiedemann and Franz in the year 1853 \cite{Ref21}. They stated that the ratio of $\kappa_e$ and $\sigma$ is proportional to temperature.      
\begin{equation}\frac{\kappa_e}{\sigma T}=L_0 \label{eq2}\end{equation}
where $L_0$ is a material- and temperature-independent constant, $L_0=\frac{\pi^2}{3}\frac{k_B}{e}^2=2.44\times10^{-8}W \Omega K^{-2}$, with $k_B$ is the Boltzmann constant and e as the electronic charge\cite{Ref21,Ref22,Ref23,Ref24}. The value of $L_0$ was first measured by Lorentz in 1872 \cite{Ref22}. Hence the name was given. Soon after that Drude attempted to measure this value theoretically. Then after the development of quantum mechanics it was tried again. And the theoretical value matched with the experimental value. Unless one takes into account inelastic collisions, metals at room temperature typically abide by this law.  Even some strongly correlated systems, such as high $T_C$ superconductors, obey the Wiedemann-Franz law \cite{Ref25} even though there are evidences of vortex excitation below the onset temperature \cite{Ref26,33,34}. However, recent studies have reported violations of this law in graphene \cite{Ref27,Ref28} . \par
Another crucial parameter to determine the thermoelectric efficiency of a material is the ZT factor. It is defined as,  $ZT=\frac{S^2\sigma T}{\kappa_e+\kappa_{ph}}$ , where total thermal conductivity $\kappa_{tot}$ has both carriers (electron and holes) and phonon contribution $(\kappa_{ph})$ such that  $\kappa_{tot}=\ \ \kappa_e+\kappa_{ph}$. Experimentally measured thermoelectric figure of merit is typically in the range of 0.01 to 0.1\cite{Ref29} This ZT factor can be affected by several parameters like substrate impurity, carrier density, etc \cite{Ref29,Ref30}. To enhance the efficiency of thermoelectric devices, one must either increase the power factor $S^2\sigma $ or decrease the $\kappa_\text{tot}$. Achieving a high ZT requires simultaneously increasing the Seebeck coefficient and electrical conductance while reducing thermal conductance, a challenging task due to the interdependence of these factors.\par
 To investigate Wiedemann Franz’s law in graphene we defined $\kappa_e/\sigma=L/L_0$.\ The validity of the law leads to  $L/L_0=1$, i.e. $L=L_0$.
Traditionally, materials like bismuth and its alloys are used in thermoelectric applications, but they are toxic, expensive, and have limited availability. A promising approach to improving ZT in new materials is leveraging the reduced phonon thermal conductance $\kappa_{ph}$ found in low-dimensional materials.\par
The paper is organized in the following way. We first describe Boltzmann transport formalism and explicitly find the scattering time $\tau$  expressions for short-range, long-range scattering, and acoustic phonon scattering for monolayer graphene \cite{Ref31,Ref32}. Next, we describe the temperature dependence of the chemical potential of a single graphene layer. We first numerically solved the equation for all temperature ranges and later derived the low and high-temperature analytical results and did comparisons. In the next section, we discuss the analytical calculations of Sommerfeld expansion. We numerically evaluate the transport coefficients and discuss and compare them with the Sommerfeld expansions in the Results and Analysis section. We conclude the paper with a summary of our results and discussions. We add two appendix sections. In Appendix 1, the detailing of Sommerfeld expansion of the Fermi integral is given.  The scaled expressions for various thermoelectric coefficients are provided in Appendix 2.
\section*{Model: Boltzmann Transport Formalism}
 In the context of linear-response approximation for electrical phenomena, the current densities are linearly dependent on the driving forces of charge transport \cite{Ref4,Ref5,Ref6}. For a thermoelectric material, those driving forces are external electric field $\left(E\right)$, gradient in number density $\left(\nabla_rn\right)$ and temperature gradient $\nabla T$. The first two forces can be combined, i.e.\ $E+\frac{1}{e}\frac{\partial \mu}{\partial n}\nabla n= \zeta$.  Where $\mu$ is the temperature-dependent chemical potential. The relation between current densities and driving forces is,
\begin{equation} \left[\begin{matrix}J\\J_Q\\\end{matrix}\right]=\left[\begin{matrix}L_{11}&L_{12}\\L_{21}&L_{22}\\\end{matrix}\right]\left[\begin{matrix}\zeta\\-\nabla\mathrm{T}\\\end{matrix}\right] \label{eq3}     \end{equation}
For diffusive cases, the transport coefficient $(L_{ij})$ can be calculated from the general expression \cite{Ref6, Ref7}
\begin{equation}\left(\begin{matrix}J_e\\J_Q\\\end{matrix}\right)=\int{\frac{d^2k}{2\pi}\left(\begin{matrix}e\\\left(E-\mu\right)\\\end{matrix}\right)v_k}\ g_k \label{eq4}\end{equation}
Here $g_{k }$ is the k-dependent nonequilibrium particle distribution function which describes the evolution of particle distribution in time under external perturbations. Using Boltzmann transport theory under relaxation time approximation, we can derive the expression for $g_{k}$. Another fundamental criterion is that the distribution has to maintain equilibrium long before and after the external perturbation is applied. So, 
\begin{equation} g_k=\tau_k\left(\frac{\partial f_0}{\partial k}\right)v_k\cdot\left(e\zeta+\frac{E_k-\mu}{T}\nabla T\right)\label{eq5}\end{equation}
Substituting this expression in Eq. (4) and converting it into the energy integral for an isotropic case we get,
\begin{equation}I_{\left(l\right)}=\int_{-\infty}^{\infty}\tau(E)\left(E-\mu\right)^{(l)}\left(-\frac{\partial f_0}{\partial E}\right)\mathfrak{D}(E)v_{F}^{2}\mathrm{d} E \label{eq6}\end{equation}
where $ \mathfrak{D}\left(E\right)=g\left|E\right|/2\pi\hbar^2v_F^2$ is the energy-dependent density of states. In this context, $E_k=\hbar v_F\left|k\right|,\ \ k$ represents the momentum, where $v_F$ is the Fermi velocity. $f_k^0$ denotes the Fermi distribution function at equilibrium. $\tau(E)$ stands for the energy-dependent relaxation time, which is proportional to the exponent of the energy, i.e. $\tau\propto\left|E\right|^m$, where $m$ is an integer representing different scattering mechanisms, and $g=g_s g_v$ corresponds to the overall degeneracy. Here, $g_s=2$  and $g_v=2$ are the degeneracies arising from spin and valley considerations, respectively.

The transport coefficients can be derived from integral $I_{\left(l\right)}$ [Eq. \ref{eq6}] as, 
\begin{equation}\  L_{11}=I_0,\ \ \ \ L_{12}=-\frac{1}{eT}I_1,\ \ L_{21}=-eI_1,\ \ L_{22}=-\frac{1}{T}I_2 \label{eq7}\end{equation}

From the integrals, the different thermoelectric parameters can be calculated as follows,

$\mathrm{Thermopower}\ S=\frac{L_{12}}{L_{11}}$, $\mathrm{Lorenz\ number\ }L=\frac{\kappa}{\sigma\ T}$, where thermal conductivity $\kappa $ and electrical conductivity  $\sigma=L_{11},\ \ \kappa_e=L_{22}+L_{21}\left(L_{11}\right)^{-1}L_{12}$. 
The scattering rate from Born approximation can be written as \cite{Ref9}
\begin{equation}\frac{1}{\tau_k}=N_i\sum_{\textbf{k}^\prime}{\frac{2\ \pi}{\hbar}\left|\left\langle\textbf{k}\middle| V_{sc}\middle|\textbf{k}^\prime\right\rangle\right|^2\delta\left(E_{\mathbf k}-E_{\mathbf{k}^\prime}\right)(1-\cos{\theta_{\textbf{k}\textbf{k}^\prime}}})\label{eq8}\end{equation}
where $N_i$ and $\theta_{\textbf{k}\textbf{k}^\prime}$ are the number of impurites and the angle between $\bf{k}$ and $\bf{k}'$ respectively, then the transition rate from the quantum state $\bf{k}$ to $\bf{k}’$ is approximated by Fermi’s golden rule. Here, $V_{sc}$ is the corresponding scattering potential due to impurities which leads the scattering of electrons from $\textbf{k}$ to $\bf{k}^{\prime}$. Under this approximation it is assumed that final states are empty and distribution function is not disturbed long after the scattering. \newline
After transforming the sum of Eq.(\ref{eq8}) into integral, it can be written as,
\begin{equation}\frac{1}{\tau_k}=\frac{n_i}{4\hbar^2 v_F}\int\frac{d^2k^{\prime}}{4\pi^2}\delta(\mathbf{k^{\prime}-k})\int d\theta_{\mathbf{k k^\prime}}\left|\langle \mathbf {k^{\prime}}\left|V_{sc}\right|\mathbf{k}\rangle\right|^2\left(1-\cos{\theta_{\mathbf{k k}^{\prime}}}\right)\label{eq9}\end{equation}
Where ,
\begin{equation*} \langle \mathbf{k}\left|V_{sc}\right|\mathbf {k}^{\prime}\rangle= \int\Psi^{*}_{\mathbf{k}^{\prime}}(\mathbf {r})V_{sc}(\mathbf{r})\Psi_{k}(\mathbf{r})d\mathbf{r}\end{equation*}
with $\Psi_{\mathbf {k}}(\mathbf{r})$ be the electronic spinor wave function and it is given by, 
\begin{equation*}\Psi_{\mathbf{k}}(\mathbf{r})=\frac{1}{\sqrt{2}}\begin{pmatrix}e^{i\theta_{\mathbf{k k^\prime}}} \\ e^{-i\theta_{\mathbf{k k^\prime}}} \end{pmatrix}e^{i \mathbf {k\cdot r}}\end{equation*}
\subsubsection*{Short range scattering:}
For short-range impurity, where the potential is given by $V_{SC}\left(\mathbf{r}\right)=V_0\delta\left(\mathbf{r}\right)$. Substituting this form of $V_{sc}$ into Eq. (\ref{eq9}) gives, 
\begin{equation}  \tau\left(E\right)=\frac{8\hbar}{N_i\pi V_0^2}\frac{1}{\mathfrak{D}\left(E\right)}\label{eq10}\end{equation}
From Eq. (\ref{eq10}) it can be easily seen that for short-range impurity $\tau\propto\left|E\right|^{-1}$
\subsubsection*{Unscreened Coloumb scattering}
For unscreened coloumb scattering $V_{sc}\mathbf{(r)}$ turns out to be the coloumb potential and it is given by,
\begin{equation*}V_{sc}(\mathbf r)=\frac{1}{4\pi\epsilon\epsilon_0}\frac{eQ}{|\mathbf{r}|}\end{equation*} 
The substrate's permitivity is $\epsilon$, while the free space permitivity is $\epsilon_0$. The electronic charge and charge of the impurity are denoted by e and Q, respectively.y. Using this potential it can be shown from the Eq.(\ref{eq9}) that,
\begin{equation}\tau(E)=\frac{16\hbar(\epsilon\epsilon_0)^2}{e^2Q^2n_{i}^{2}}|E| \end{equation}
\subsubsection*{Screened Coloumb scattering}
Let's now consider the impact of the long-range Coulomb potential. Charged impurities are located in the insulating $SiO_2$ layer and are screened by the conduction electrons in the graphene sheet. Consequently, the potential in momentum space can be described as follows:
\begin{equation}\phi\left(q\right)=\frac{1}{\epsilon\epsilon_0q}\rho\left(q\right)+\frac{Ze}{\epsilon\epsilon_0}\frac{1}{q}e^{-q\left|a_c\right|}\label{eq11}\end{equation}
where $\rho(q)$ represents the induced charge density, and $a_c$ indicates the shortest distance from the externally charged impurity to the two-dimensional graphene sheet.\par
Since we are using a semiclassical approach, it is consistent to approximate the induced charge density within the Thomas-Fermi (TF) approach \cite{Ref32}\newline
Under TF approach,
\begin{equation*}\rho\left(\mathbf{r}\right)\approx e\sum_{\mathbf{k}}\left[f\left(E_\mathbf{k}-e\phi\left(\mathbf{r}\right)\right)-f\left(E_\mathbf{k}\right)\right)=e\phi\left(\mathbf{r}\right)\sum_{\mathbf{k}}\left(-\frac{\partial f_\mathbf{k}^0}{\partial E_k}\right)\end{equation*}
This finally gives, 
\begin{equation}\rho\left(\mathbf{r}\right)=-e^2\phi\left(\mathbf{r}\right)\mathfrak{D}\left(E_F\right)\label{eq12}\end{equation}
Here it is assumed that Fermi surface is “sharp” for this system.
\newline
So, under this TF approach the screening is modified in the following way,
\begin{equation*}\phi\left(q\right)=\frac{Ze}{2\epsilon_0\epsilon}\frac{e^{-q\left|a_c\right|}}{q+\gamma}\end{equation*}
where $\gamma=\mathfrak{D}\left(E_F\right)e^2/2\epsilon_0\epsilon$.
'Using Eq. (\ref{eq8}), $V_{SC}=e\phi\left(q\right)$ and $a_c=0$ for large doping
\begin{equation}\tau_k=\frac{\hbar v_Fk_F}{u_0^2}    \text{With the} \  u_0=\frac{\sqrt{N_i}Ze^2}{4\epsilon_0\epsilon\left(k_F+\gamma\right)}\label{eq13}\end{equation}
From Eq. (\ref{eq13}) it is established that the scattering time is energy-independent for long-range screened Coulomb scattering.
\section*{Analytical calculations}
\subsection*{Temperature dependence of chemical potential}
In this section, we aim to numerically investigate the temperature dependence of the chemical potential. To provide readers with insight into this behavior, we will derive asymptotic expressions for the chemical potential at very low and very high temperatures. These derivations will utilize the Sommerfeld expansion and the Riemann zeta function, respectively.\par
The carrier density (N) is the excess carrier concentration when the lower band is completely filled and N=0 while the chemical potential $\mu\left(T\right)=0$. At any fixed temperature N remains constant and the expression is given by
\begin{equation}
N=\int_{0}^{\infty}f\left(E\right)\mathcal{D}\left(E\right)dE-\int_{-\infty}^{0}\left(1-f\left(E\right)\right)\mathcal{D}\left(E\right)dE\label{eq14}
\end{equation}
where $f(E)=\frac{1}{1+\exp{\left(E-\mu(T)\right)}/k_BT}$ is the Fermi-Dirac distribution function. At zero  temperature T=0 K, all the states up to the Fermi energy $E_F=\mu(0)$ are completely filled and the remaining states are empty and therefore $f\left(E\right)=1 \text{for} \  E\leµ(0) \ \text{and}  \ f\left(E\right)=0$  for  $E>µ(0)$. Using this fact, the carrier density can be obtained as, \newline
$N=\int_{0}^{E_F}{\mathcal{D}\left(E\right)dE=\alpha\ \frac{E_F^2}{2}}$ where  $\alpha=\frac{g_sg_v}{2\pi\left(\hbar\ v_F\right)^2}=1.5\times{10}^8\frac{cm^{-2}}{meV^2}$. \newline 
$\mu\left(T\right)$ is a temperature-dependent quantity and at $T=0, \mu_0=E_F=k_BT_F$. To determine the temperature dependence $\mu\left(T\right)$ we first rewrite Eq. (14) after suitable scaling as the following expression,
\begin{equation}
\frac{1}{2}\left(\frac{T_F}{T}\right)^2=-\left(\mathrm{Polylog}\left(2,-\exp{\left(-\frac{\mu(T)}{k_BT}\right)}\right)+\mathrm{Polylog}\ \left(2,-\exp{\left(\frac{\mu(T)}{k_BT}\right)}\right)\right)\label{eq15} 
\end{equation}
Here Polylog is the polylogarithmic function. It is defined as,
                           $\mathrm{Polylog}\left[\nu,-z\right]=\Gamma(\nu)f_{\nu}\left(z\right)$   where $f_\nu\left(z\right)$ is Fermi-Dirac integral [Eq.  (\ref{eq16})]                                                                                  
The chemical potential is obtained by fixing the number density N in Eq. (\ref{eq14}). We numerically solve Eq.  (\ref{eq15}). We first found a set of roots $\frac{\mu\left(T\right)}{k_B\ T}$  of the right hand side expression of Eq. (\ref{eq15}) for a set of values of $\frac{T}{T_F}$ and suitably multiplied by a factor of  $\frac{T}{T_F}$ to the set of roots (which was interpolated to generate more number of roots) $\frac{\mu\left(T\right)}{k_B\ T}$ to finally obtain $\frac{\mu\left(T\right)}{k_B\ T_F}$ as a function of scaled temperature $\frac{T}{T_F}$. The numerical result is shown by the green curve of Fig 2. The numerically evaluated result is compared with the two asymptotic expressions obtained from low temperature expansions and high temperature expansions respectively. The exact matching of the with the analytical and numerical results in the two temperature regimes validate the correctness of the numerical evaluation of  $\frac{\mu\left(T\right)}{k_B\ T_F}$. The two analytical curves approach very close to each other at around $T\approx0.43-0.54\ T_F$ 
\begin{figure}[htp]\centering \includegraphics[width=3.5in,height=2in]{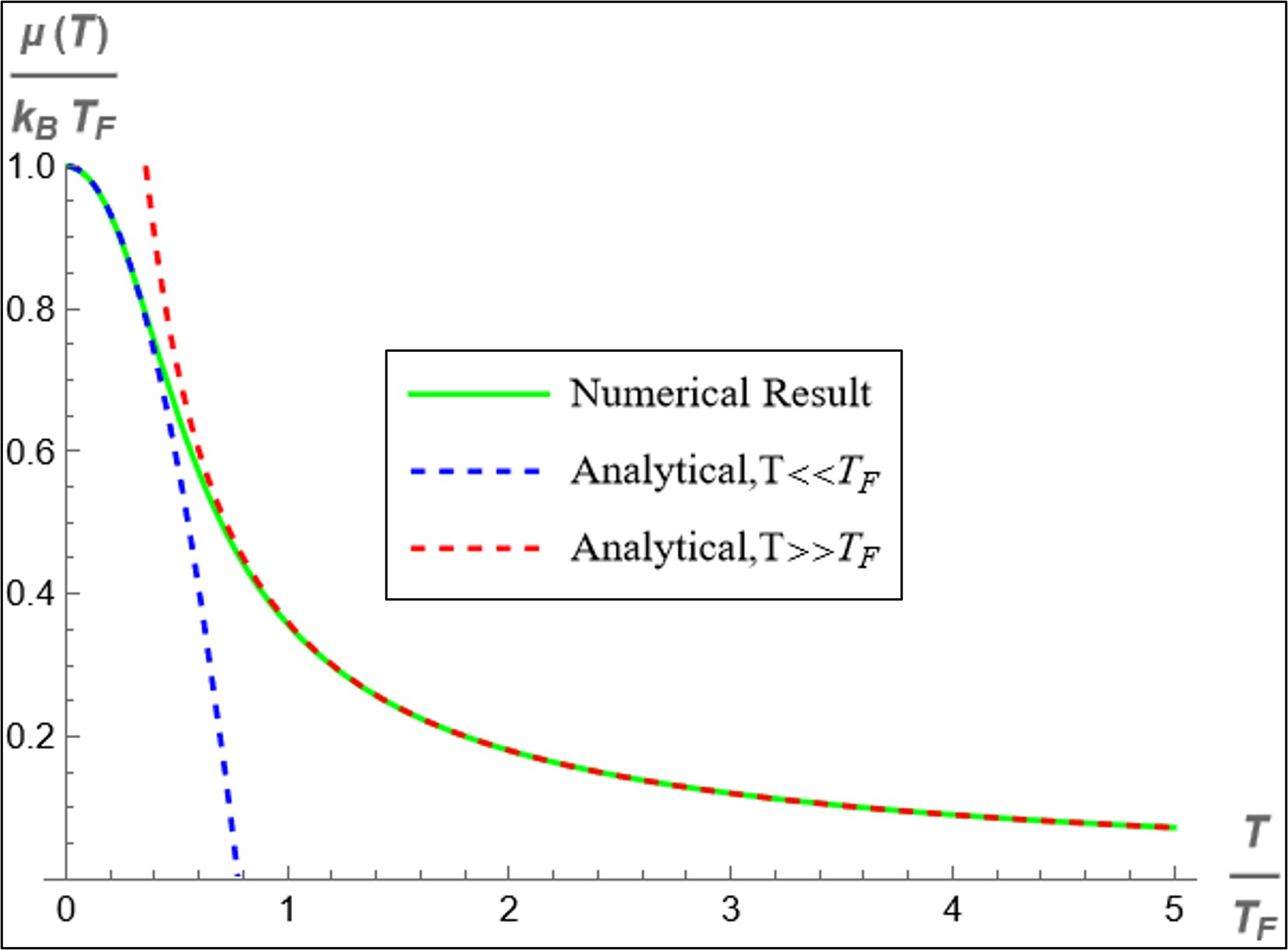}\caption{Scaled chemical potential $\frac{\mu\left(T\right)}{k_B\ T_F}$ is plotted as a function of scaled temperature $\frac{T}{T_F}$. The solid line is the numerically evaluated result. The blue dashed line represents the low-temperature asymptotic result given by Eq.  (\ref{eq19}) and the red line is the high-temperature asymptotic result given by Eq. (\ref{eq25}) }\label{fig2}\end{figure}\par\
To derive the analytical expressions we first recall the Fermi-Dirac integral, which is defined as,
\begin{equation}f_\nu\left(z\right)=\frac{1}{\Gamma\left(\nu\right)}\int_{0}^{\infty}\frac{x^{\nu-1}}{1+z^{-1}e^x}\ dx \label{eq16}\end{equation}
where, $z=e^\frac{\mu}{k_BT}$, known as the fugacity factor. Now if $T$ is very low, i.e. $T\ll T_F$, Fermi-Dirac integrals can be written as (See Appendix A1 for more details and also we define $\alpha=ln\ z=\frac{\mu}{k_BT}$
\begin{equation}f_\nu\left(z\right)=\frac{\alpha^\nu}{\Gamma\left(\nu+1\right)}\left[1+\nu\left(\nu-1\right)\frac{\pi^2}{6}\frac{1}{\alpha^2}+\nu\left(\nu-1\right)\left(\nu-2\right)\left(\nu-3\right)\frac{7\pi^4}{360}\frac{1}{\alpha^4}\cdot\cdot\right]  \label{eq17}\end{equation}
\subsection*{Low temperature}
Considering there would not be enough thermal energy for holes to occupy higher energy states than the Fermi level at low enough temperatures, we may ignore the hole contribution from the right side of Eq. (\ref{eq14}) at low temperatures. Then substituting Eq. (\ref{eq16}) into Eq. (\ref{eq14}), 
\begin{equation*}\frac{\left(k_BT_F\right)^2}{2}=2\left(k_BT\right)^2f_2\left(z\right)\end{equation*}
Using Eq. (\ref{eq16}),
\begin{equation*}\frac{\left(k_BT_F\right)^2}{4}=\frac{\mu^2}{\Gamma\left(2+1\right)}\left[1+2\left(2-1\right)\frac{\pi^2}{6}\frac{\left(k_BT\right)^2}{\mu^2}+\cdot\cdot\cdot\cdot\right]\end{equation*}
 We obtain the low-temperature expression of the chemical potential as,
\begin{equation}        \mu=E_F\left(1-\frac{\pi^2}{6}\frac{T^2}{T_F^2}\right)  \ \ \ \text{for} \ \ \ T<<T_F     \label{eq18}      \end{equation}
\subsection*{High Temperature}
Since in our current consideration, $T\gg T_F$ and $z\approx1$, we can retain the first few terms only. Under this approximation, the Fermi-Dirac integral represented by Eq. (\ref{eq16}) can be expressed as,
\begin{equation}f_\nu\left(z\right)=z-\frac{z^2}{2^2}+\frac{z^3}{3^2}-\cdot\cdot  \label{eq19}\end{equation}
 After making a suitable substitution $( x=\frac{E}{k_BT})$ on the right-hand side of Eq. (\ref{eq13}) or Eq. (\ref{eq14})
\begin{equation}\left(k_BT\right)^2\left[\int_{0}^{\infty}{\frac{x}{1+z_1^{-1}e^x}dx}-\int_{-\infty}^{0}{\frac{x}{1+z_2^{-1}\ e^x}dx}\right]  \label{eq20}\end{equation}

For the 1st integral of Eq. (\ref{eq21}), we substitute the expression directly from Eq. (\ref{eq20}) and we obtain
\begin{equation}\left(1+\frac{\mu}{k_BT}\cdot\cdot\right)-\left(\frac{1}{4}+\frac{\mu}{2k_BT}\cdot\cdot\right)+\left(\frac{1}{9}+\frac{\mu}{3k_BT}\cdot\cdot\right) \label{eq21}\end{equation}
For the 2nd integral of Eq. (\ref{eq21}) where $z_2=e^{-\ \frac{\mu}{ k_BT}}$ under variable transformation,$ x\rightarrow-x^\prime$  the above equation will be modified as,
\begin{equation}\left(1-\frac{\mu}{k_BT}\cdot\cdot\right)-\left(\frac{1}{4}-\frac{\mu}{2k_BT}\cdot\cdot\right)+\left(\frac{1}{9}-\frac{\mu}{3k_BT}\cdot\cdot\right) \label{eq22}\end{equation}
By substitution Eq. (20) and Eq. (21) into Eq. (19) we get, 
\begin{equation}\frac{\left(k_BT_F\right)^2}{2}=2\frac{\mu}{k_BT}\left(1-\frac{1}{2}+\frac{1}{3}\cdot\cdot\cdot\right) \label{eq23}\end{equation}
This directly gives the high temperature expression of the chemical potential $\left(\mu\right)$ as,
\begin{equation}\frac{\mu}{k_BT_F}=\frac{1}{4\log{2}}\frac{T_F}{T} \ \ T\gg T_F \label{eq24}\end{equation}
\subsection*{Sommerfeld expansion of thermoelectric coefficients}
\subsubsection*{Electrical conductivity: }
The electrical conductivity is defined earlier as  $\sigma=L_{11}$. Using Eq. (\ref{eq18}), we can determine the low-temperature expression for electrical conductivity. It can be shown that the electrical conductivity can be expressed as,
\begin{equation}\sigma=\int_{0}^{\infty}\mathcal{D}\left(E\right)\left(-\frac{\partial f_0}{\partial E}\right)dE=\alpha\tau_0\left(m+1\right)\int_{0}^{\infty}\frac{E^mdE}{1+z^{-1}e^{\left(\frac{E}{k_BT}\right)}}  \label{eq25}\end{equation}
Using Eq. (26) after suitable substitution, we get,
\begin{equation}\sigma=\alpha\tau_0\left(k_BT\right)^{m+1}\left(m+1\right)\Gamma\left(m+1\right)f_{m+1}\left(z\right)=\alpha\tau_0\ \mu^{m+1}(1+\left(m+1\right)\frac{\pi^2}{6}\left(k_BT\right)^2\mu^{-2}) \label{eq26}\end{equation}
        Eq. (\ref{eq16}) represents the low-temperature behavior of electrical conductivity.
\subsubsection*{Seebeck coefficient}
The Seebeck coefficient is defined as $S=\frac{L_{12}}{L_{11}}$. Using the same method now we derive a low-temperature expression for the transport coefficient $L_{12}$.                      
\begin{align}(-{eT)\ L}_{12}&=\alpha\tau_0\int_{0}^{\infty}{\left(E-\mu\right)\left|E\right|^{m+1}\left(-\frac{\partial f_0}{\partial E}\right)dE} =\alpha\tau_0\left(m+2\right)\int_{0}^{\infty}\frac{E^{m+1}dE}{1+z^{-1}e^\frac{E}{k_BT}}-\mu\left(m+1\right)\int_{0}^{\infty}\frac{E^mdE}{1+z^{-1}e^\frac{E}{k_BT}}\nonumber\\&\ =\alpha\tau_0\left(m+2\right)\int_{0}^{\infty}\frac{E^{m+1}dE}{1+z^{-1}e^\frac{E}{k_BT}}-\mu\left(m+1\right)\int_{0}^{\infty}\frac{E^mdE}{1+z^{-1}e^\frac{E}{k_BT}}\nonumber\\&=\ \alpha\tau_0\ \mu^{m+2}\left[\left(1+\left(m+2\right)\left(m+1\right)\frac{\pi^2}{6}\frac{\left(k_BT\right)^2}{\mu^2}\right)-\left(1+\left(m+1\right)\left(m\right)\frac{\pi^2}{6}\frac{\left(k_BT\right)^2}{\mu^2}\right)\right]\nonumber\\&= \alpha\tau_0\ \mu^{m+2}\left[\left(1+\left(m+2\right)\left(m+1\right)\frac{\pi^2}{6}\frac{\left(k_BT\right)^2}{\mu^2}\right)-\left(1+\left(m+1\right)\left(m\right)\frac{\pi^2}{6}\frac{\left(k_BT\right)^2}{\mu^2}\right)\right]  \label{eq27}\end{align}
So from the expression of $L_{12}$ and $L_{11}$ S can be calculated as,
\begin{equation} S=-\frac{1}{eT}\frac{\mu\left(\left(m+1\right)m\frac{\pi^2}{3}\frac{\left(k_BT\right)^2}{\mu^2}\right)}{1+\left(m+1\right)\frac{\pi^2}{6}\left(k_BT\right)^2\mu^{-2}}  \label{eq28}\end{equation}
Under low temperatures, the quadratic term of the denominator is very less than 1. So it can be neglected safely. Also, at $T\ll T_F$, $\mu$ can be approximated as the Fermi energy $E_F$. Using this approximation we can write, 
\begin{equation}S=-\frac{k_B}{e}\frac{\pi^2}{3}\frac{T}{T_F} \label{eq29} \end{equation}
\subsubsection*{Thermal conductivity:}
Now we should aim to derive a similar expression for thermal conductivity. The thermal conductivity $\left(\kappa_e\right)$  can be expressed in terms of transport coefficients as,
\begin{equation}\kappa_e=-L_{22}+\frac{L_{12}L_{21}}{L_{11}} \label{eq30}\end{equation}
So, we try to derive an expression for $L_{22}$ under Sommerfeld expansion, from BTE,
\begin{align}  \left(-T\right)\ L_{22}&=\int_{0}^{\infty}{\left(E-\mu\right)^2\left(-\frac{\partial f_0}{\partial E}\right)\mathcal{D}\left(E\right)}\ =\alpha\tau_0\int_{0}^{\infty}{\left(E-\mu\right)^2\left(-\frac{\partial f_0}{\partial E}\right)E^{m+1}dE}\nonumber\\&=\alpha\tau_0\int_{0}^{\infty}\left(\left(m+1\right)E^m+2\left(E-\mu\right)E^{m+1}\right)f\left(E\right)dE=\alpha\tau_0\int_{0}^{\infty}\left(\left(m+1\right)E^m+2\left(E-\mu\right)E^{m+1}\right)f\left(E\right)dE\nonumber\\&=\alpha\tau_0\int_{0}^{\infty}\left[\left(m+3\right)E^{m+2}-2\mu\left(m+2\right)E^{m+1}+\mu^2\left(m+1\right)E^{m+1}\right]f\left(E\right)dE\nonumber\\&=\alpha\tau_0\ \left(k_BT\right)^{m+3}\left(m+3\right)\int_{0}^{\infty}{\frac{x^{m+2}}{1+z^{-1}e^x}dx}-2\mu\left(k_BT\right)^{m+2} \alpha\tau_0\left(m+2\right)\int_{0}^{\infty}{\frac{x^{m+1}}{1+z^{-1}e^x}dx}\nonumber\\&  +\mu^2\left(k_BT\right)^{m+2}\left(m+2\right)\int_{0}^{\infty}{\frac{x^{m+1}}{1+z^{-1}e^x}dx}\nonumber\\&=\ \alpha\tau_0\left(k_BT\right)^{m+3}\Gamma\left(m+4\right)f_{m+3}\left(z\right)-2\mu\left(k_BT\right)^{m+2}\Gamma\left(m+3\right)f_{m+2}\left(z\right) +{ \mu}^2\left(k_BT\right)^{m+2}\Gamma\left(2\right)f_{m+1}\left(z\right)  \label{eq31}      
 \end{align}
upon further simplification, Eq. (\ref{eq31}) becomes,
\begin{equation}L_{22}=-\alpha\tau_0\ \mu^{m+3}\frac{\pi^2}{3}\left(\frac{k_BT}{\mu}\right)^2  \label{eq32}\end{equation}
Using Onsager’s symmetry relation, $L_{12}=TL_{21}$ and therefore substituting the respective expressions (derived above) in Eq. (\ref{eq29}) we get the expression of 
\begin{equation} \kappa_e=\frac{1}{T}\mu^{m+3}\frac{\pi^2}{3}\left(\frac{k_BT}{\mu}\right)^2\left[1+\left(\left(m+1\right)^2\frac{\pi^2}{3}\left(\frac{k_BT}{\mu}\right)\right)^2\right]  \label{eq33} \end{equation}
Since we have already derived the low-temperature expression of both $\kappa_e$ and $\sigma$, we can also give a low-temperature expression for the Lorentz number. The analysis shows that,
\begin{equation} \frac{L}{L_0}=\left(1+\left(m+1\right)^2\frac{\pi^2}{3}\left(\frac{T}{T_F}\right)^2\right) \label{eq34}\end{equation}
\subsubsection*{Internal energy \& specific heat :}
The total internal energy can be calculated as, 
\begin{align}  U\left(T\right)=\int_{0}^{\infty}E\mathcal{D}\left(E\right)f\left(E\right)\ dE=\alpha\tau_0\int_{0}^{\infty}\frac{E\ dE}{1+e^\frac{E-\mu}{k_BT}} \label{eq35} \end{align}
The Sommerfeld expansion for the Eq. (\ref{eq23}) can be written as,
\begin{equation}U\left(T\right)=\alpha\tau_0\left(k_BT\right)^3\Gamma(3)f_3\left(z\right)=\frac{1}{3}\alpha\tau_0\left(\mu\right)^3\left(1+\pi^2\left(\frac{\mu}{k_BT}\right)^2\right)  \label{eq36}\end{equation}
At $T\ll T_F$,  $\mu\approx k_BT_F$ then,
\begin{equation}U\left(T\right)=\frac{1}{3}\alpha\tau_0\left(1+\pi^2\left(\frac{T}{T_F}\right)^2\right)  \label{eq37}\end{equation}
The numerically evaluated internal energy by  Eq. (\ref{eq35})  and the corresponding low-temperature expression of U given by Eq.(\ref{eq36}) are plotted in Figure (4).
Again, from the definition of specific heat, 
\begin{equation}  C_V=\frac{\partial U}{\partial T} \label{eq38}\end{equation}
\begin{equation}N=\int D\left(E\right)f\left(E\right)dE=\alpha\tau_0{{(k}_BT)}^2\Gamma(2)\ f_2\ \left(z\right)  \label{eq39}\end{equation}
so,
\begin{equation}C_V=2k_B\frac{\partial\left({(k}_BT)\frac{f_3\left(z\right)}{f_2\left(z\right)}\right)}{\partial T}=2k_B\left(3\frac{f_3\left(z\right)}{f_2\left(z\right)}-2\frac{f_2\left(z\right)}{f_1\left(z\right)}\right) \label{eq40}\end{equation}
Further simplification of Eq. (\ref{eq40})  gives,
\begin{equation}C_V=C_0\frac{T}{T_F} \ \ \text{where} \ \ C_0=\frac{2\pi^2}{9}k_B  \label{eq41} \end{equation}

\section*{Numerical Results and Analysis}
This section is devoted to the evaluation of numerical results and their analysis. We exclusively considered diffusive charge transport and calculated various thermoelectric coefficients numerically. Before doing the numerical calculation we first write down all the transport coefficients in terms of convenient and dimensionless forms, the details are given in the above sectionPlease look at that section to get all the expressions.  \par
\begin{figure}[ht]
\centering
\includegraphics[width=4in]{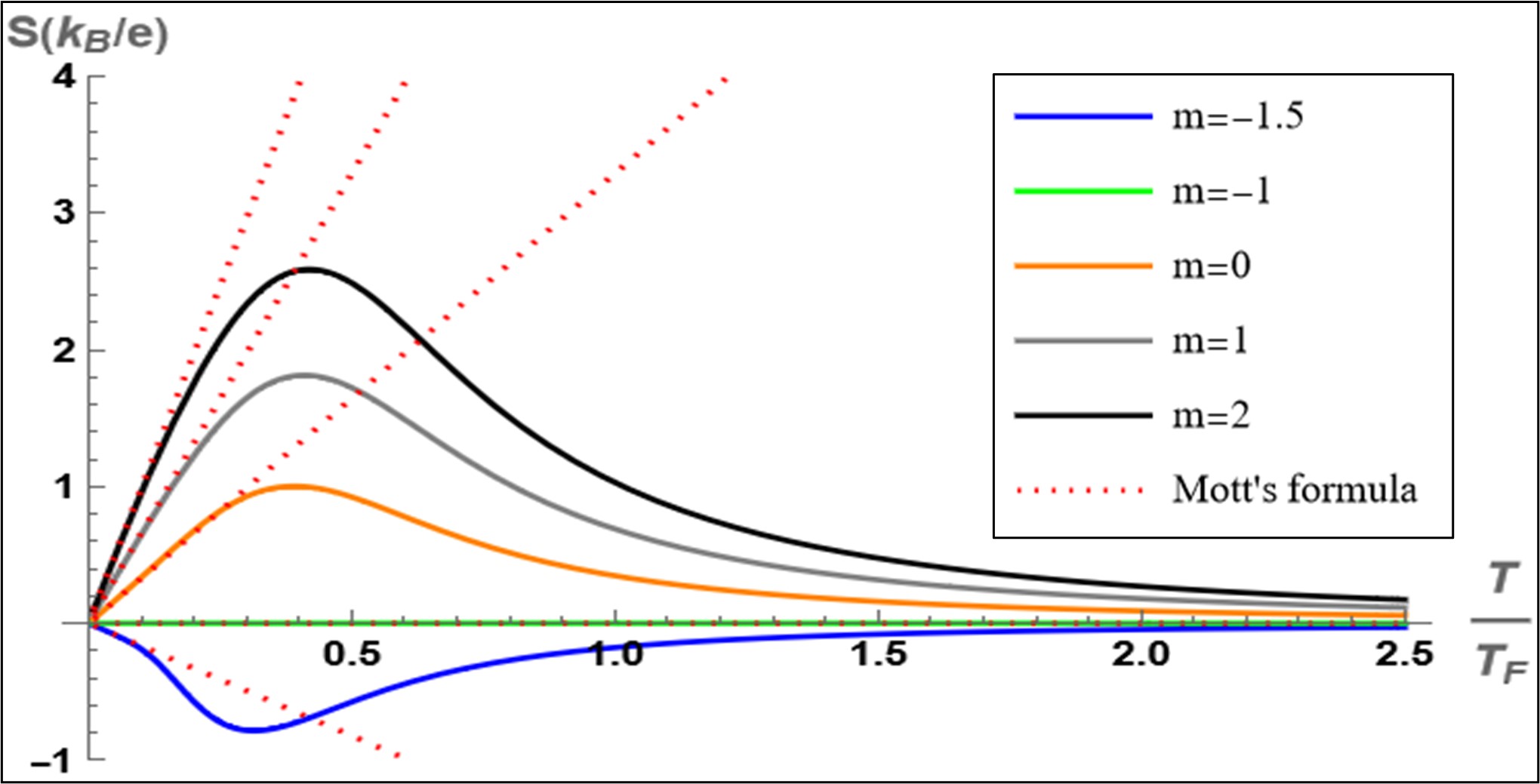}
\caption{ Seebeck coefficient in units of $k_B/e$ is plotted against scaled temperature $T/T_F$  for different scattering mechanisms related to m (solid lines). Along with this low-temperature analytical expression is plotted, which is derived using Sommerfeld expansion of the Fermi integral (dotted lines)..}
\label{fig3}.
\end{figure}
We first show the numerically evaluated thermopower (S) result using Eq.(\ref{eq54} ) in Fig.(\ref{fig3}) where the variation of S (in units of $k _B/e$) is plotted as a function of $T/T_F$  And compared it with the low temperature (left) and high temperature analytical results (right). Note that, the value of 1  $k_B/e =86 \mu V/K$. Our numerically obtained values are consistent with the experimentally obtained results \cite{Ref5} of a single layer of graphene on a substrate having a maximum value of S in the range of $80-100 \mu V/K$. Interestingly our numerically obtained results are consistent with Mott’s formula (Eq. \ref{eq29}) in the low-temperature regime for all values of m, which is compared by dotted red lines in Fig. 3. The plot shows graphene has a high value of thermopower in the range $0$ to $1.2 T_F$ for all values of m>1. Note that the thermopower peak height increases as m increases.  For m< -1 thermopower shows a sign change (positive to negative) and at these values of m negative peak is observed. At low temperatures i.e.  $T<<T_F$ those thermopower curve approaches Mott’s law approximately at   $T\leq0.25T_F$   for $m>1$ and for $m<1$ the matching is even less.\par
\begin{figure}[ht]\centering 
\includegraphics[width=4in,height=2.5in]{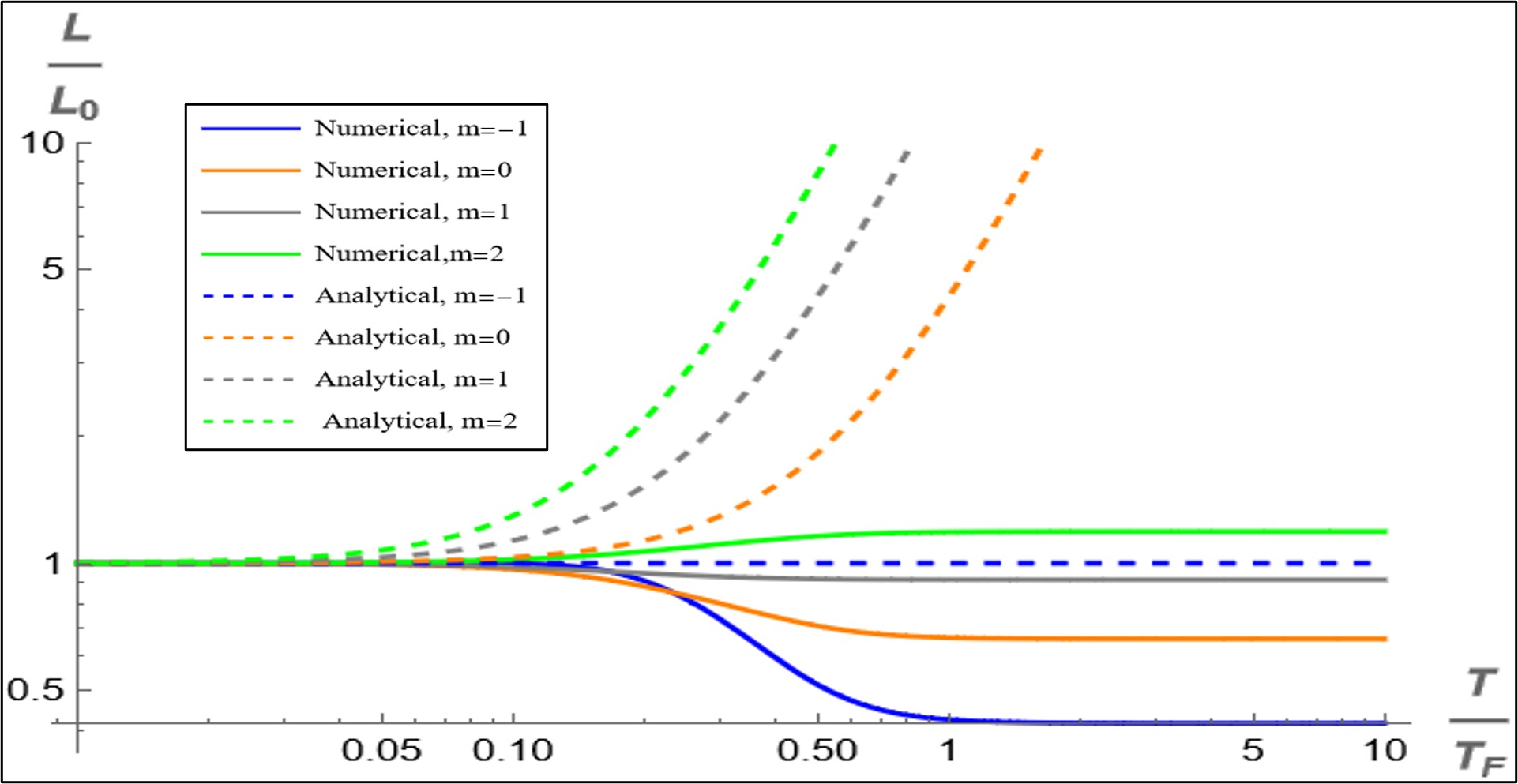}
\caption{Lorentz number behavior is shown under temperature variation for different scattering. The variation of $L/L_0$ at $T<<T_F$ is shown by dotted lines. (b) Comparision of analalytical high temperature Lorentz number with the numerical results. }\label{fig4}
\end{figure}
We now directly show the results for Lorentz number $L/L_0$ calculations as a function of $T/T_F$. In Fig. \ref{fig4}. Here, both the axes are labeled on logarithmic scales. The numerical results using Eq. (\ref{eq56}) are also compared with the low-temperature analytical calculations using Sommerfeld expansions using  Eq. (\ref{eq34}) . While $T/T_F \leq0.2$, the numerical and analytical curves are approaching 1 for all scattering parameters m. As T approaches 0 K, Wiedemann Franz’s law must be obeyed for any finite carrier density consistent with our numerical and low-temperature analytical results. As the temperature is raised from $0.13T_F$ to $T_F$ for all values of m, the numerical values $L/L_0$ are either decreasing from 1 (for m = -1, 0 and 1) or remain almost constant (m = 1) while the Sommerfeld results are increasing in this range, where high-temperature analytical calculations need to be performed to match with numerical results, which shows the Lorentz number curves are flat, indicating saturation of the Lorentz number. The saturation value depends on m. Therefore, a modified m-dependent Wiedemann-Franz’s law can be established for $T>T_F$. In a recent work \cite{Ref27}, the experimentally determined value of $L/L_0$ showed these kinds of behavior at very low and very high-temperature regimes, as mentioned in our Fig. 4, but a peculiar peak (see Fig. 3(b) of Ref. [27]) was observed in addition at around T = 60 K, where the value of $L/L_0$ was ~20 in the cleanest sample. Such a large value is attributed to the presence of Dirac fluid near the charge neutral point. Such high values of $L/L_0$ are also attempted to explain by developing the bipolar diffusive carrier transport model in the literature \cite{Ref28} and have recently gained significant attention in the research community. \par 
\begin{figure}[ht] 
\includegraphics[width=3.5in,height=2.5in]{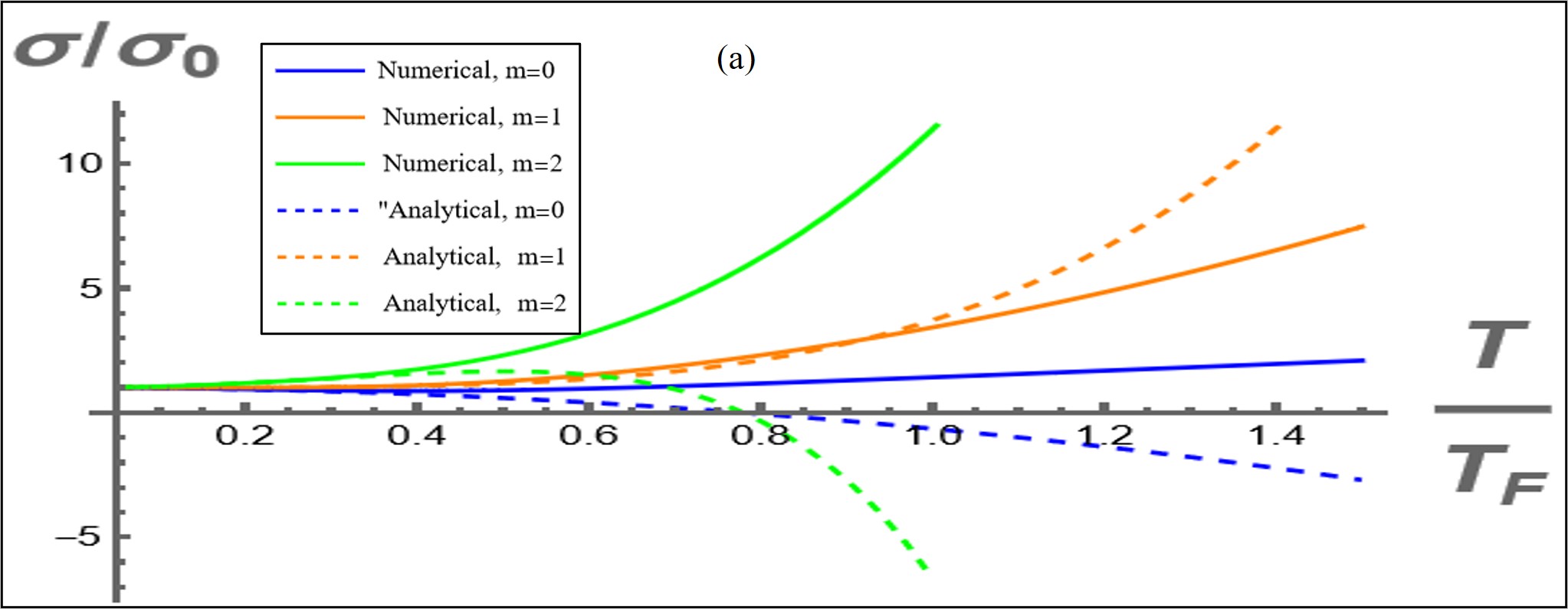} 
 \includegraphics[width=3.5in,height=2.5in]{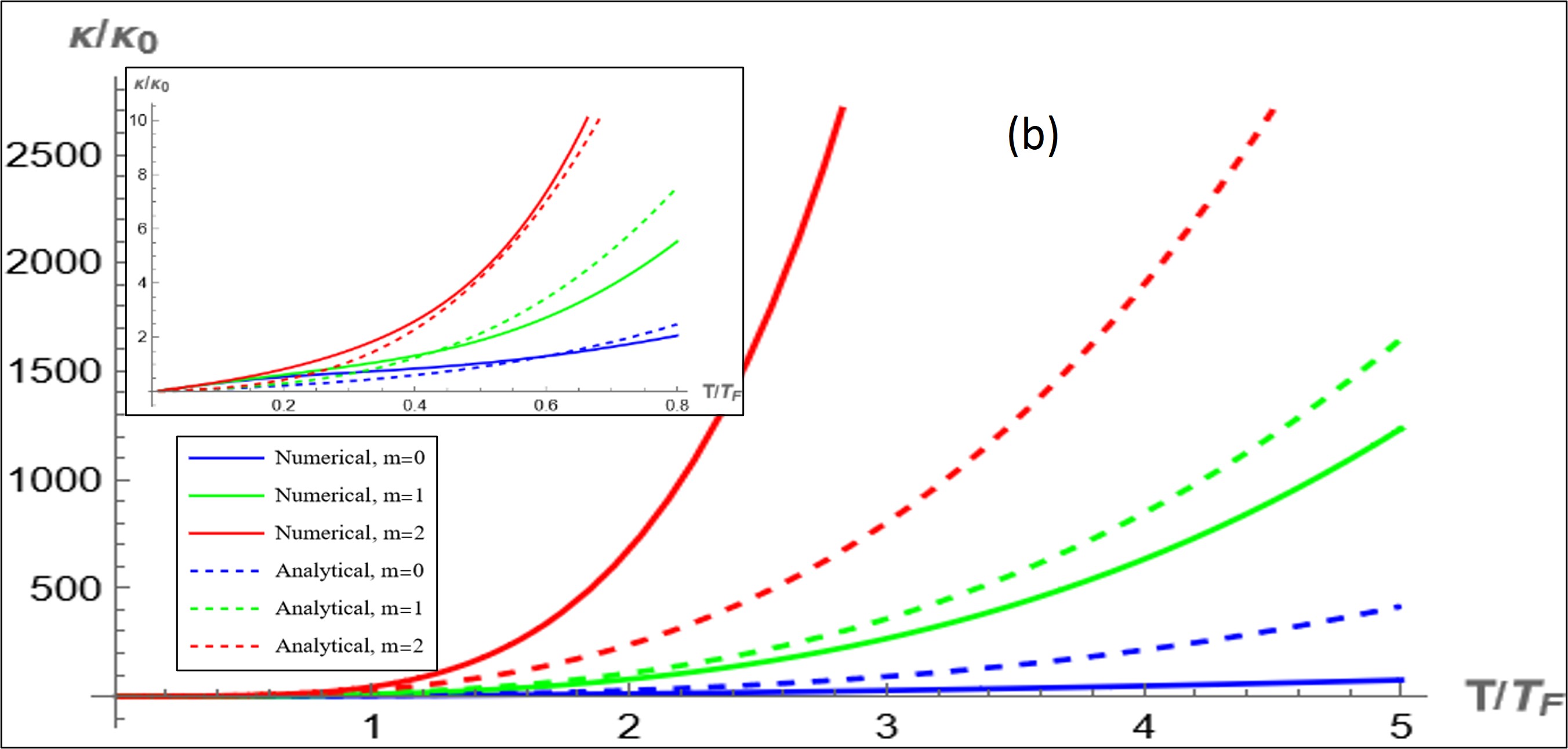}
 \caption{(a) Depiction of electrical and (b) thermal conductivity (electronic) variation against temperature scaled by Fermi temperature $\left(T_F\right)$. The dotted line represents the analytical expression for $\sigma$ and $\kappa_e$ derived from Sommerfeld expansion.  Inset: The region where $\frac{\kappa}{\kappa_0}$ is linear as function of temperature }\label{fig5}\end{figure}
Fig. \ref{fig5}(a), represnts the electrical conductivity variation against the temperature. Here, the solid lines indicate the numerical results, and the dotted line represents the analytical expression given in Eq. (\ref{eq27}). The numerical evaluation is done using the expression Eq. (\ref{eq53}) The graph shows electrical conductivity has some minimum vaule even at zero temperature. The experimentally determined value of $\sigma$ is $1-1.5 S/m^2$  at room temperature \cite{Ref15}. In our case $\sigma$ is scaled by $\sigma_0$. The value of $\sigma_0$ is in the order of 1 S/m. The curve for  m=0 which corresponds to the scattering due to screened coloumb potential due to the substrate impurity shows the $\sigma$ has the value around 1-2 $\sigma_0$. Then as the temperature increases $\sigma$ increases monotonously. The anlytical and numerical plots match well only at $T<T_F$. $\sigma$ is increasing more rapidly after $T_F$ with the increasing values of m. \par
 On the other hand, In Fig. \ref{fig5}(b) thermal conductivity is plotted against temperature for different scattering mechanisms with Eq. (\ref{eq55}). The curve tells us that $\kappa$ is increasing linearly. As m increases the linear nature of the $\kappa_e$ decreases. Here the dotted line represents the low temperature behavior which is derived analytically and plotted using Eq. (\ref{eq33}). The picture shows that analytical and numerical results have good agreement up to $0.7T_F$. Then the deviation occurs. \par

Typically, experimentally measured thermal conductivity ranges from 3500 to 5000 W/mK at room temperature \cite{Ref18}.  However, in our plot, the values range from 50 to 100 $\kappa_0$, where $\kappa_0$ is determined by the expression $\frac{\left(k_BT_F\right)^{m+4}}{2\pi\hbar^2v_F^2e^2}$. Here, we calculated only the electronic contribution to $\kappa_{tot}$, and $\tau_0$ is approximately ${10}^{-13}$s ($l/v_F$, where l is the mean free path of the electron.).  The calculated value of $\kappa$ is much lower than experimental value because we have considered only the instead of $\kappa_{tot}$. And $\kappa_e$ contributes only less than 1 \% of the $\kappa_{tot}$ \cite{Ref20}
\begin{figure}[htp]\centering 
\includegraphics[width=4in,height=2.2in]{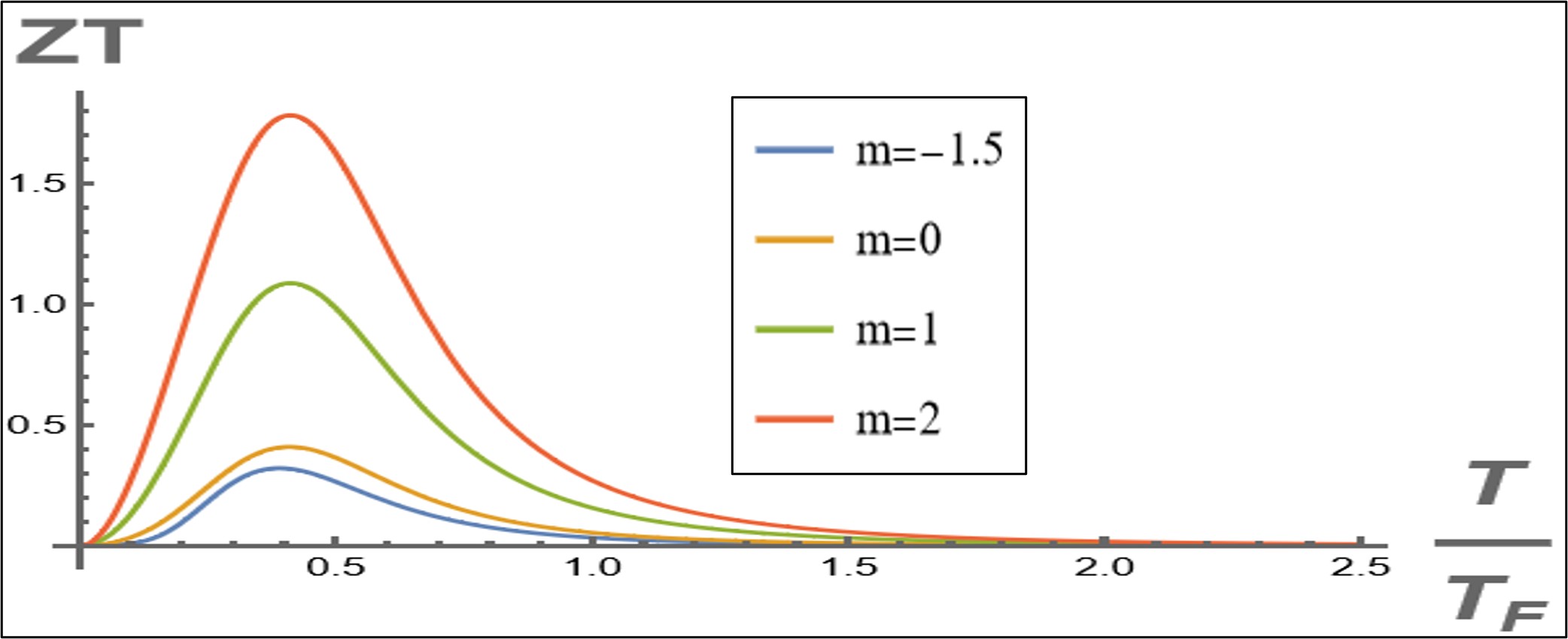}
\caption{ Thermoelectric figure of merit (ZT factor) is plotted against the variation of temperature. }\label{fig6}
\end{figure}
For short-range impurity and in-plane acoustic phonon scattering, the relaxation time depends on energy as $\tau=\tau_0\left|E\right|^{-1}$. As a result, the integral $L_{12}$ is zero upon full integration, leading to a zero thermopower and, correspondingly, a zero ZT factor over the entire temperature range.
Conversely, for long-range Coulomb scattering where m=0, the thermopower shows a significant peak up to the Fermi temperature, with a peak height of 0.41 at $0.4T_F$. This peak height increases significantly with increasing values of m. For $m\geq1$, the results show that the ZT factor exceeds 1.\par
Typically, the thermoelectric figure of merit for single-layer graphene is between 0.01 and 0.1 at room temperature \cite{Ref29}. While comparing with the experimental result there would be deviation as the contribution phonon contribution of thermal conductivity is not incorporated in our Boltzmann transport calculations. The ZT factor would be reduced further if the phononic contribution were taken into account. In recent times the power factor has increased so that the thermoelectric figure of Merit comes up to 0.4\cite{Ref29}. However, the reduction of $\kappa_{ph}$ is still challenging. \par
In Figure \ref{fig6}(a) the variation of total internal energy is shown. Here we have only considered the electronic specific heat instead of total specific heat It is proportional to the thermal conductivity and the exact relation between $\kappa_e$ and $C_V$ is,
\begin{equation*}\kappa_e=\frac{1}{3}\lambda\ v_FC_V\end{equation*}
Where $\lambda$ is the mean free path and $v_F$ is the Fermi velocity of the electrons. The variation of electronic specific heat is shown in Fig. \ref{fig6}(b)
\begin{figure}[htp] 
\includegraphics[width=3.4in]{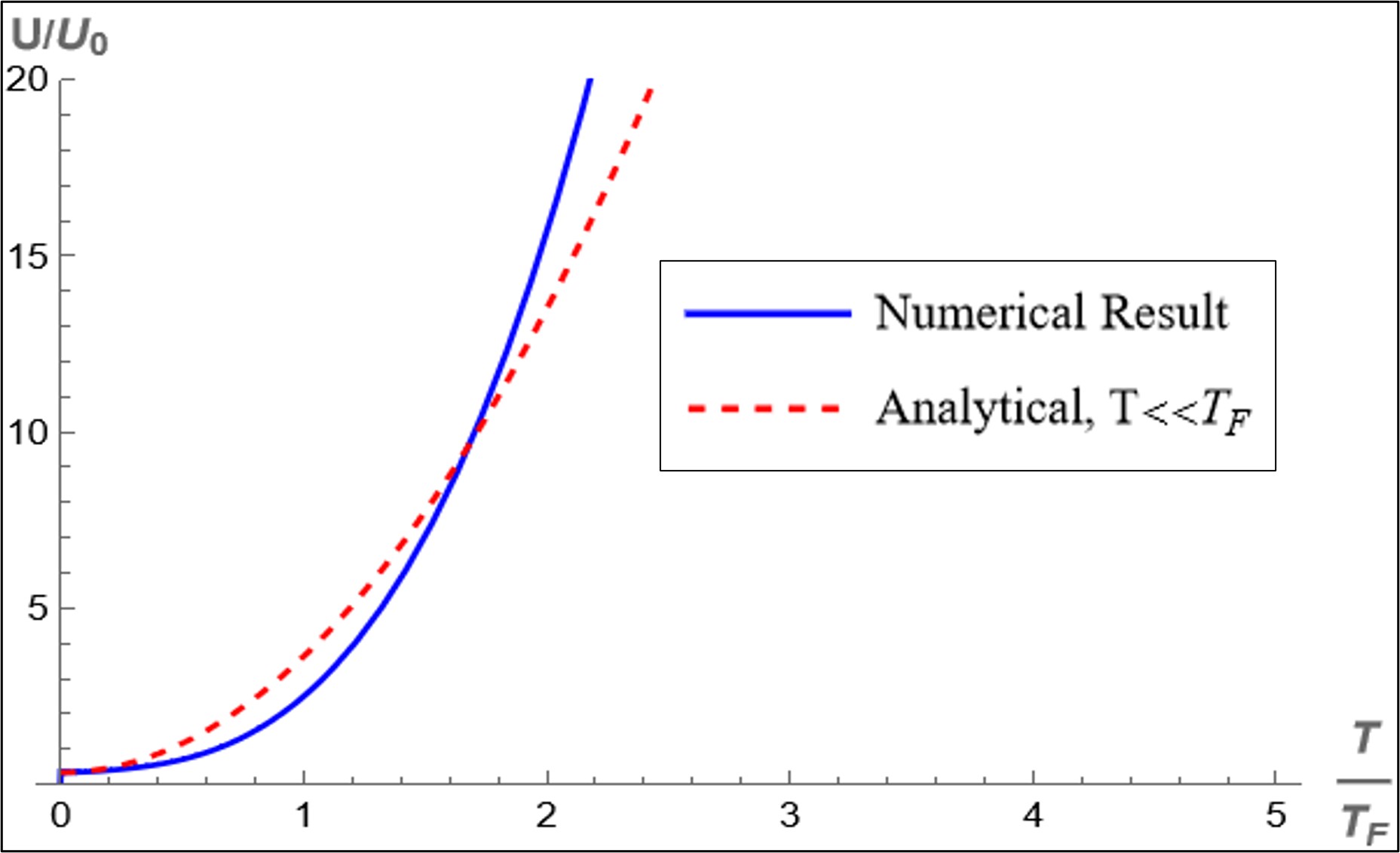}\hspace{0.2cm}
\includegraphics[width=3.4in]{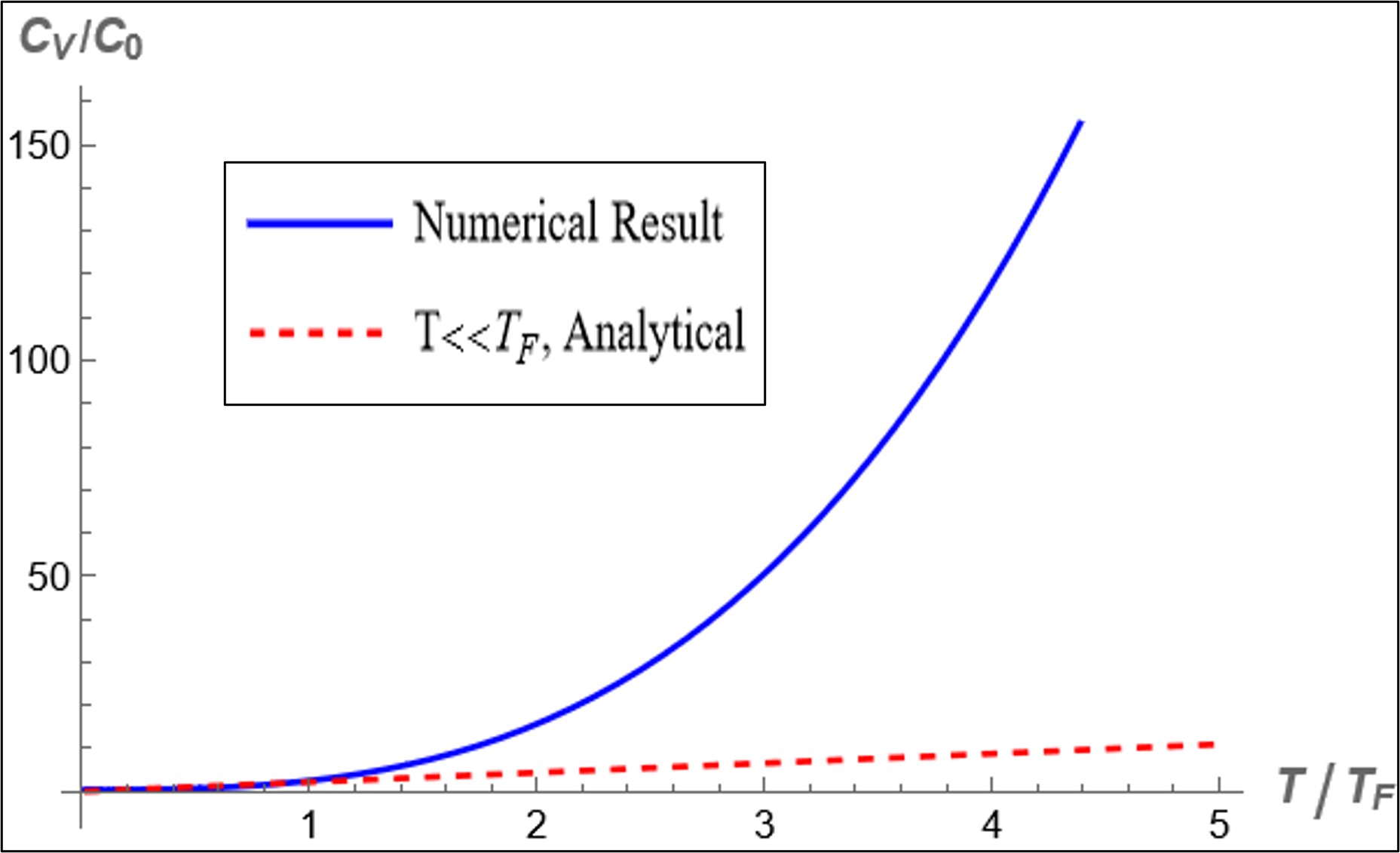}
\caption{(a) Plot of total internal energy and (b) specific heat against scaled temperature. The corresponding low temperature analytical results has also been depicted by the dashed line.}\label{fig7}\end{figure}

\section*{Conclusion}In conclusion, we have studied the thermoelectric transport properties of graphene. In this theoretical analysis, we explored the validity of Mott’s and Wiedemann-Franz’s law under different scattering mechanisms. Our theoretical and numerical calculations confirm Mott’s law for  $T\ll T_F$. Beyond this temperature range, deviations from Mott’s law are observed for every scattering mechanism except for $m= -1$. A sign change in thermopower occurs when $m< -1$. We also showed numerical and analytical results for the Lorentz number and compared them with the existing experimental and theoretical results. Saturation of the Seebeck coefficient (S) and Lorentz number L are seen for all values of m. More numerical results on electronic and thermal conductivity are provided and compared with the existing results in the literature. The thermoelectric figure of merit, or ZT factor, calculations are shown here, with a peak in the temperature range of 0 to $T_F$. Our calculations show promising agreement with numerical analysis. An analytical expression is derived for electrical and thermal conductivity using the Sommerfeld expansion of the Fermi integral. We aimed to offer a detailed review of this material, providing valuable insights that could make this nanomaterial promising for future thermoelectric applications, such as addressing energy waste due to Joule heating in nanomaterials, microfiber electronics, etc.   
\section*{Acknowledgment}
 The authors would like to thank for technical and software support of high performance computing (HPC) lab, VIT-AP University for providing computational resources. K.S. acknowledges the support of VIT-AP University Research Grant in Engineering Management and Sciences (RGEMS) grant with order number VIT-AP/SpoRIC/RGEMS/2024-2025/005.

\section*{Appendix 1: Derivation of Sommerfeld expansion}\label{appnd1}
Here we try to derive the Eq. $\left(\ref{eq16}\right)$ for Fermi integral. For our purpose, $z\gg1$, we introduce a variable,
\begin{align}   \alpha=ln\left(z\right)  \label{eq42}\end{align}
Let’s define two functions,
\begin{equation}F_\nu\left(e^\alpha\right)=f_\nu\left(e^\alpha\right)\Gamma\left(\nu\right)=\int_{0}^{\infty}\frac{x^{\nu-1}}{e^{x-\alpha}+1}dx  \label{eq43} \end{equation}
for large values of $\alpha$, the dominating factor in that function is $\frac{1}{e^{x-\alpha}+1}$. Its' departure from the limiting value is significant only at the close proximity of $x=\alpha$. So, the simplest approximate function is a step function as a result, the integral in Eq. (\ref{eq43}) is reduced to,
\begin{equation}F_\nu\left(\alpha\right)\approx\int_{0}^{\alpha}x^{\nu-1}dx=\frac{\alpha^\nu}{\nu} \label{eq44}\end{equation}
correspondingly
\begin{equation} f_\nu\left(e^\alpha\right)\approx\frac{\alpha^\nu}{\Gamma\left(\nu+1\right)}  \label{eq45}\end{equation}
For better approximation, we can write Eq. $\left(\ref{eq43}\right)$ as
\begin{equation}\int_{0}^{\alpha}{x^{\nu-1}\left[1-\frac{1}{e^{\left(\alpha-x\right)+1}}\right]dx}+\int_{\alpha}^{\infty}\frac{x^{\nu-1}dx}{e^{x-\alpha}+1} \label{eq46}\end{equation}
here we do the substitution $x=\alpha-\eta_1 \text{and} x=\alpha+\eta_2$ in the respective integrals after this substitution we get,
\begin{equation}F_\nu\left(\alpha\right)=\frac{\alpha^\nu}{\nu}-\int_{0}^{\alpha}\frac{\left(\alpha-\eta_1\right)^{\nu-1}d\eta_1}{e^{\eta_1}+1}+\int_{0}^{\infty}\frac{\left(\alpha+\eta_2\right)^{\nu-1}d\eta_2}{e^{\eta_2}+1} \label{eq47}\end{equation}
 since $\alpha\gg1$ the upper limit of the integral may be safely replaced by $\infty$. Also $\eta_1=\eta_2=\eta$, again,
\begin{equation}F_\nu\left(\alpha\right)=\frac{\alpha^\nu}{\nu}-\int_{0}^{\infty}\frac{\left(\alpha+\eta\right)^{\nu-1}-\left(\alpha-\eta\right)^{\nu-1}}{e^\eta+1}d\eta  \label{eq48}\end{equation}
After expanding the numerator binomially we get,
\begin{equation}F_\nu\left(\alpha\right)=\ \frac{\alpha^\nu}{\nu}+2\sum_{k=1,3,5,\cdot\cdot} C_k^{\left(\nu-1\right)}\alpha^{\nu-k-1}\int_{0}^{\infty}{\frac{\eta^k}{e^\eta+1}\ d\eta} \label{eq49}\end{equation}
\section*{Appendix 2: Scaling of transport coefficients}\label{appnd2}
For numerical evaluation (we have used the Wolfram Mathematica software system for generating all the above figures) of chemical potential and all other thermoelectric transport coefficients, we introduce three dimensionless quantities $x,\ y,\ \text{and} t$ to convert the energy E, chemical potential µ and temperature T in a dimensionless form such that $x=\frac{E}{k_BT_F},\ \ y=\frac{\mu}{k_BT_F} \text{and},\ t=\ \ \frac{T}{T_F}$. Using these dimensionless quantities, the expression $L_{ij}$ turns out to be 
\begin{equation}L_{11}=\frac{g\tau_0\left(k_BT_F\right)^{m+2}}{2\pi\hbar^2v_F^2T_Ft}\int_{-\infty}^{\infty}{dx\frac{\exp{\left(\frac{x-y}{t}\right)}}{\left(1+\exp{(}\frac{x-y}{t})\right)^2}\left|x\right|^{m+1}}  \label{eq50}\end{equation}
\begin{equation}L_{12}=\frac{1}{e}\frac{g\tau_0\left(k_BT_F\right)^{m+3}}{2\pi\hbar^2v_F^2T_F^2t^2}\int_{-\infty}^{\infty}dx\frac{\exp{\left(\frac{x-y}{t}\right)}}{\left(1+\exp{(}\frac{x-y}{t})\right)^2}\left(x-y\right)\left|x\right|^{m+1} \label{eq51}\end{equation}
\begin{equation}L_{22}=-\frac{g\tau_0\left(k_BT_F\right)^{m+4}}{2\pi\hbar^2v_F^2T_F^2t^2e^2}\int_{-\infty}^{\infty}dx\frac{\exp{\left(\frac{x-y}{t}\right)}}{\left(1+\exp{(}\frac{x-y}{t})\right)^2}\left(x-y\right)\left|x\right|^{m+1} \label{eq52}\end{equation}

The electrical conductivity expression in terms of  x, y, and t transforms into
\begin{equation}  \frac{\sigma}{\sigma_0}=\int_{-\infty}^{\infty}dx\frac{\exp{\left(\frac{x-y}{t}\right)}}{\left(1+\exp{(}\frac{x-y}{t})\right)^2}\left(x-y\right)\left|x\right|^{m+1} \label{eq53}\end{equation}
 where $\sigma_0=\ \frac{g\tau_0\left(k_BT_F\right)^{m+2}}{2\pi\hbar^2v_F^2T_F}$ .  Similarly, the thermopower $S=\frac{L_{12}}{L_{11}}$  reduces to
\begin{equation}S=\frac{k_B}{e}\frac{T_F}{T}\frac{\int_{-\infty}^{\infty}dx\frac{\exp{\left(\frac{x-y}{t}\right)}}{\left(1+\exp{\left(\frac{x-y}{t}\right)}\right)^2}\left(x-y\right)\left|x\right|^{m+1}}{\int_{-\infty}^{\infty}{dx\frac{\exp{\left(\frac{x-y}{t}\right)}}{\left(1+\exp{(}\frac{x-y}{t})\right)^2}\left|x\right|^{m+1}}} \label{eq54}\end{equation}
Finally, the thermal conductivity $\kappa$ expression reduces to
\begin{equation}   \frac{\kappa}{\kappa_0}=\left(-L_{22}+\frac{L_{12}L_{21}}{L_{11}}\right)\frac{1}{t}   \  ,\text{ where} \    \kappa_0= -\frac{g\tau_0\left(k_BT_F\right)^{m+4}}{2\pi\hbar^2v_F^2e^2}  \label{eq55}          \end{equation}
where   $\kappa_0= -\frac{g\tau_0\left(k_BT_F\right)^{m+4}}{2\pi\hbar^2v_F^2e^2}$
Lorentz number L is calculated as, 
\begin{equation}\frac{L}{L_0}=\frac{1}{t}\frac{\left(\frac{\left(\int_{-\infty}^{\infty}dx\frac{\exp{\left(\frac{x-y}{t}\right)}}{\left(1+\exp{\left(\frac{x-y}{t}\right)}\right)^2}\left(x-y\right)\left|x\right|^{m+1}\right)^2}{\int_{-\infty}^{\infty}{dx\frac{\exp{\left(\frac{x-y}{t}\right)}}{\left(1+\exp{(}\frac{x-y}{t})\right)^2}\left|x\right|^{m+1}}}\right)+\int_{-\infty}^{\infty}dx\frac{\exp{\left(\frac{x-y}{t}\right)}}{\left(1+\exp{(}\frac{x-y}{t})\right)^2}\left(x-y\right)\left|x\right|^{m+1}\ \ \ \ }{\int_{-\infty}^{\infty}dx\frac{\exp{\left(\frac{x-y}{t}\right)}}{\left(1+\exp{(}\frac{x-y}{t})\right)^2}\left(x-y\right)\left|x\right|^{m+1}} \label{eq56}\end{equation}

\end{document}